# Dark Energy, Paradigm Shifts, and the Role of Evidence

Ofer Lahav (Department of Physics & Astronomy, University College London) and Michela Massimi (School of Philosophy, Psychology and Language Sciences, University of Edinburgh).
E-mails: o.lahav@ucl.ac.uk; michela.massimi@ed.ac.uk

**Abstract:** We comment on cases in the history of Astronomy, which may shed some light on the current established but enigmatic concordance model of Cosmology. Should the model be understood by adding new entities such as Dark Matter and Dark Energy, or by modifying the underlying theory? For example, the prediction and discovery of planet Neptune can be regarded as analogous to finding a dark component; while explaining the anomalous perihelion precession of Mercury by General Relativity can be taken as analogous to the possibility that modified gravity is an alternative to dark components of the universe. In this paper, we revise this analogy coming from the history of astronomy with an eye to illustrating some of the similarities and differences between the two cases.

**Dark Energy: the status quo**

Cosmological measurements, recently confirmed and refined by the Planck space mission and other probes, strongly favor a 'concordance' model, according to which the universe is flat and contains approximately 5% ordinary matter (baryons), 25% Cold Dark Matter and 70% Dark Energy (see the Planck Collaboration 2013 study and references therein). The concept of Dark Energy is a variant on Einstein's Cosmological Constant, Lambda ($\Lambda$), and the proposition for a $\Lambda$-like linear force can even be traced back to Newton (e.g. Calder & Lahav 2008, 2010 for a historical perspective). This '$\Lambda$ + Cold Dark Matter' ($\Lambda$CDM) paradigm and its extensions pose fundamental questions about the origins of the Universe. If Dark Matter and Dark Energy truly exist, we must understand their nature. Alternatively, General Relativity and related assumptions may need radical modifications. These topics were flagged as key problems by researchers and by advisory panels around the world. Commonly, Dark Energy is quantified by an equation of state parameter, $w$, which is the ratio of pressure to density. The case $w = -1$ corresponds to Einstein's cosmological constant in General Relativity, but in principle $w$ may vary with cosmic epoch, e.g. in the case of scalar fields. Essentially, $w$ affects both the geometry of the universe and the growth rate of structures. These effects can be observed via a range of cosmological probes, including the Cosmic Microwave Background (CMB), galaxy clustering, clusters of galaxies, and weak gravitational lensing, in addition to Supernovae Ia. The Hubble diagram of Type Ia Supernova (Perlmutter et al. 1999; Riess et al. 1998), for which the 2011 Nobel Prize in Physics was awarded, revealed that our universe not only is expanding but is also accelerating in its expansion. The main problem is that we still have no clue as to what is causing the acceleration, and what Dark Matter and Dark Energy are.

The key point we are addressing in this article is the following: should a discrepancy between data and the existing cosmological theory be resolved by adding new entities such as Dark Matter and Dark Energy, or by modifying the underlying theory? The Dark Energy Survey (see Box 1) and other similar projects aim to address this important question by looking for further experimental evidence for Dark Energy.

There is still a possibility for another major paradigm shift in our understanding of the cosmos, including the following options:

*(i) Violation of the Copernican Principle*: for example, if we happen to be living in the middle of a large void;

*(ii) Dark Energy being something different than vacuum energy:* although vacuum energy is mathematically equivalent to $\Lambda$, the value predicted by fundamental theory is as much as $10^{120}$ times larger than observations permit;

*(iii) Modifications to gravity*: it may be that General Relativity requires revision to a more complete theory of gravity;

*(iv) Multiverse*: if $\Lambda$ is large and positive, it would have prevented gravity from forming large galaxies, and life would never have emerged. Using this Anthropic reasoning to explain the Cosmological Constant problems suggests an infinite number of universes ('Multiverse') in which $\Lambda$ and other cosmological parameters take on all possible values. We happen to live in one of the universes, which is 'habitable'.

---

**Box 1: The Dark Energy Survey (DES, http://decam.fnal.gov)**
DES is using a new wide-field camera for the Blanco 4m telescope in Chile (see Figures 1 and 2). By 2018 DES will map 200 million galaxies over 1/8 of the sky, in 5 optical filters, and it will detect thousands of Supernovae Type Ia. The survey had its First Light in September 2012 and started survey observations in September 2013. It will run over 525 nights spread over five years. The main goal of DES is to characterise the Dark Energy and other key cosmological parameters to high precision. DES will measure these using four complementary techniques in a single survey: counts of galaxy clusters, weak gravitational lensing, galaxy clustering and Type Ia Supernovae. DES is an international collaboration including 300 scientists from the US, the UK, Spain, Brazil, Germany and Switzerland.

There are several other projects with similar goals, some under constructions, others proposed. Imaging surveys include HSC, Pan-STARRS, KIDS, PAU and JPASS, alongside spectroscopic surveys, for example WiggleZ, BOSS, e-BOSS, DESI, HETDEX, PFS, 4MOST, SKA and Euclid.

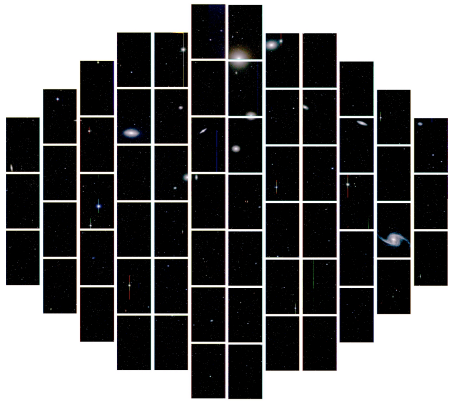

Figure 1: The patch of sky in the direction of the Fornax cluster observed by DES across it 570 Megapixel CCD mosaic. For interactice viewing of the objects see http://www.darkenergysurvey.org/dark-energy-camera-mosaic/

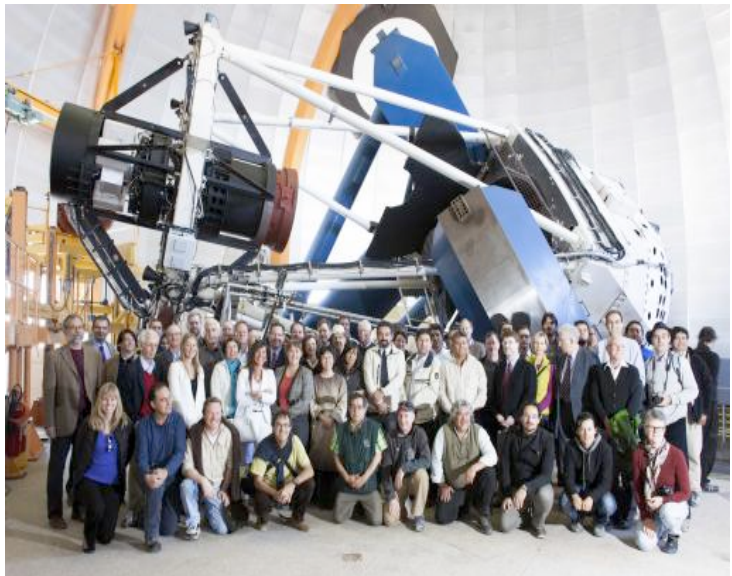

Figure 2: The Dark Energy Camera on the Blanco 4pm telescope in Chile. Photo taken at the DES inaugural ceremony in November 2012.

**"Haven't we been there already?"**
**The discovery of Neptune and the perihelion of Mercury**

While waiting for the results of DES and other surveys, one can speculate about how the future of cosmology is going to look like. Other cases in the history of Astronomy and Physics shed some light on our understanding of the current concordance model (see Table 1). A famous past episode in the history of Astronomy is instructive to this purpose. Consider the discovery of the planet Neptune predicted by Adams and Le Verrier, back in 1846. The anomalous perihelion of the planet Uranus had been known for some time. In the 1820s, the astronomer Friedrich Wilhelm Bessel hypothesize a possible departure from Newton's inverse square law of gravity to account for the anomalous perihelion of Uranus, due to specific gravity varying from one body to another. But the hypothesis was experimentally falsified and abandoned by the 1840s (for details see R. W. Smith 1989 p. 399).

Two astronomers, Urbain Le Verrier and John Couch Adams independently had tried to reconcile this piece of negative evidence with Newtonian theory by postulating the existence of a new planet, called Neptune, having a certain mass and at a certain distance from the orbit of Uranus, which could explain the anomalous perihelion (for historical details, see Grosser 1962).

The new planet was indeed observed on 23 September 1846, the actual position having been predicted with a good degree of accuracy by Adams and Le Verrier. Yet, when Le Verrier applied a similar line of reasoning for the anomaly of the planet Mercury, by postulating also in this case a new planet, called Vulcan, whose mass and orbit could explain the observed anomaly, no such planet was observed. Despite early attempts to understand the 43 arc seconds/century of Mercury anomalous perihelion by modifying Newton's inverse square law of gravity (Hall 1894 and Newcomb 1895), a final explanation of the phenomenon came only with the advent of General Relativity.

The episode of Neptune versus Vulcan has been the battleground of important methodological discussions among philosophers of science since the early twentieth century. Karl Popper famously referred to the discovery of Neptune as an attempt by Adams and Le Verrier to shelter Newton's theory from falsification by introducing the auxiliary hypothesis of a new planet (Popper, 1974, p. 986; see also Bamford 1996). For Popper, scientists should strive to falsify scientific theories by imposing severe tests on them. Theories that survive severe testing get corroborated; theories that fail to pass severe tests get falsified and rejected. But obviously, in the case of well-supported and highly successful theories, such as Newtonian mechanics, the discovery of a piece of negative evidence (such as the anomalous perihelion of Uranus) did not necessarily lead scientists to falsify and reject the theory straightaway (despite some isolated speculations). Instead, the vast majority of the scientific community at the time chose to hold on to Newtonian mechanics, and accommodate the newly discovered piece of negative evidence by adding the further assumption that there could be another planet, Neptune.

**Hard core and protective belts**

The philosopher of science Imre Lakatos used this same episode to criticize Popper's view, according to which severe tests and falsifiability are the distinctive criteria of scientific knowledge. For Lakatos (1970), the discovery of Neptune showed instead that hardly ever theories get refuted by a single piece of negative evidence, and a better way of thinking about scientific theories is in terms of what Lakatos called "research programmes". In Lakatos' terminology, a "research programme" (say, Newtonian mechanics) has two key components: a *hard core* (namely, the main theoretical assumptions; say, Newton's three laws of motion and the law of gravitation); and a *protective belt* of auxiliary assumptions (namely, additional hypotheses that are often used to protect the main theoretical assumptions from falsification; e.g., the number of planets in the solar system). Lakatos argued that any research programme comes with a *negative heuristic* that restrain scientists from attacking the hard core, by tweaking the protective belt of auxiliary assumptions instead. While the *positive heuristic* envisages ways in which a research programme may be tweaked to make novel predictions, which turn out to be correct (distinguishing in this way between what Lakatos called *progressive* research programmes versus *degenerating* ones, i.e. those whose predictions turn out to be false, leading to the demise of the research programme itself). The discovery of Neptune was—to Lakatos's eyes—an example of the progressive nature of the Newtonian research programme.

More generally, philosophers of science have used the example of Neptune vs. Vulcan as an illustration of the so-called Duhem–Quine thesis (see Gillies, 1993, 98–116). The thesis claims the following: given that scientific theories consist of both main theoretical hypotheses and auxiliary assumptions, in the case of recalcitrant evidence, it is never clear whether scientists should go for the option of modifying the auxiliary assumptions (e.g., number of planets, as in the case of Neptune), or for the alternative option of revising the main theoretical hypotheses themselves (e.g., from Newtonian mechanics to General Relativity, as in the case of the anomalous perihelion of Mercury). The French physicist and philosopher Pierre Duhem (1906/1991), who first stumbled into this peculiar feature of scientific theories, famously suggested that scientists follow their 'good sense' when facing dilemmas of this kind.

Thus, one may wonder whether the current state of the art in cosmology (for a philosophical analysis, see Smeenk 2013) resembles the discovery of Neptune, or whether the recalcitrant evidence coming from Supernova Ia may not be better explained by a modification of the accepted paradigm (like in the case of the perihelion of Mercury, which ushered in General Relativity). Going down the latter route would require modifying General Relativity itself, and/or rejecting the standard Friedman-Lemaître-Robertson-Walker (FLRW) models (as opposed to introducing Dark Energy). For Dark Matter, rejecting the current 'concordance' paradigm would imply an even more drastic revision of our accepted and well-entrenched theories of gravity.

On galactic scales it remains a mystery why rotation curves of spiral galaxies are 'flat', i.e. with constant rotation velocity out to large radius from the galaxy's centre. If the galaxy mass is concentrated at the centre like the luminous matter, one would expect the rotation velocity to decline with distance. The most common explanation, assuming Newtonian gravity, is that galaxies are embedded in massive halos, which generate the flat rotation curves. An alternative is to modify Newtonian gravity at the limit of low acceleration (Milgrom 1983, Bekenstein 2010), so that the observed flat rotation curves can be obtained without invoking Dark Matter halos. While the current paradigm is that Dark Matter halos exit, until Dark Matter is actually detected directly it would be difficult to deduce which of the two options (or others) is valid.

So do we really need to introduce new kinds of matter? In the history of Physics, this has often proved to be the case. Consider, as another example this time from nuclear physics, Wolfgang Pauli's hypothesis of the neutrino in 1933 (see Massimi 2005, ch. 4). Pauli was working within the framework of Dirac's hole theory, whereby holes in the negative energy sea were interpreted as fermionic anti-particles (i.e. positrons). He concluded that in the beta decay, it was neither possible for a proton to decay into a neutron and a positron, nor for a neutron to decay into a proton and an electron, on pain of either violating the conservation of angular momentum, or ascribing integral or null spin to the neutron. And while other physicists, such as Bohr and G.P. Thomson, entertained the hypothesis of conservation laws being violated, for angular momentum to be conserved in beta decay, Pauli hypothesized the existence of a new particle: what became later known as the neutrino. The hypothesis of the neutrino was taken on board by Fermi in his 1934 work on the beta decay, although experimental evidence for the new particle did not become available until after Cowan, Reines etal (1956) experiments with nuclear reactors.

| Phenomenon | New Entity | New theory |
| --- | --- | --- |
| Uranus' orbit | Neptune | (Bessel's specific gravity ruled out) |
| Mercury's orbit | (Hypothetical planet Vulcan ruled out) | General Relativity |
| Beta decay | Neutrino | (violation of angular momentum ruled out) |
| Galaxy flat rotation curves | Dark Matter? | Modified Newtonian Dynamics? |
| Accelerating universe (SN Ia and other data) | Dark Energy? | Modified General Relativity? |

Table1: Examples of alternatives of new entity vs. new theory.

**Paradigm shifts and the role of experimental evidence**

When does a piece of recalcitrant evidence point to a paradigm shift? And under what conditions can we hold on to the accepted paradigm, without running the risk of scientific conservatism? Fifty-two years ago, the historian and philosopher of science Thomas Kuhn published a highly influential book entitled *The Structure of Scientific Revolutions* (Kuhn, 1962). Kuhn portrayed the development of science as a cyclic sequence of normal science, crisis, and scientific revolutions, whereby an old paradigm—successful as it might have been for centuries—faces an increasing number of anomalies that force scientists to abandon it and to move on to a new paradigm. Famously, by 'paradigm' Kuhn meant to capture not only the scientific theory endorsed by a scientific community at a given time (say, Aristotelian physics or Newtonian mechanics or General Relativity), but also the experimental and technical resources available to the community at the time, the cultural influences and even the system of values embraced. Kuhn defined the passage from one paradigm to another—be it the passage from Ptolemaic astronomy to Copernican astronomy, or from Newtonian mechanics to Relativity—in terms of 'incommensurability'.

Scientific paradigms are incommensurable because (using the metaphor of the relation between the side and the diagonal of a triangle) they lack a 'common measure' to evaluate whether the later paradigm improves on the earlier one. Scientific methodology and experimental resources are so ingrained within each paradigm (and its conceptual resources) that—in Kuhn's view—it is not possible to have cross-paradigm standards and norms of scientific evaluation. Kuhn's new vision of science targeted a view dominant at the time in philosophy of science, which portrayed scientific knowledge as a cumulative and serendipitous sequence of theories, each one building on the predecessor, and more likely to be true than its predecessors.

One of the central questions that Kuhn's account left nonetheless wide open is how many anomalies are necessary to induce a paradigm shift. More to the point, when does a piece of recalcitrant evidence (like the one coming from Supernova Ia) qualify as 'anomalous', in Kuhn's sense, to induce a scientific revolution? When can it be reconciled with the accepted paradigm instead (by, say, modifying some suitable auxiliary hypothesis, as for example in the case of Pauli's neutrino, the discovery of Neptune, galactic dark halos, or the hypothesis of Dark Matter and Dark Energy)? Are we really left with Duhem's 'good sense' in grappling with these questions?

More work needs be done to answer these questions. It is only with wisdom of hindsight that cosmologists will be able to answer these pressing issues. The empirical findings of DES and other experiments will prove pivotal in this respect. In the case of Dark Matter, in addition to cosmological pieces of evidence (such as gravitational lensing), non-cosmological findings too (in the form of galactic dynamics and direct search for Dark Matter) will also be crucial. From a philosophical point of view, the overall empirical support that General Relativity has received so far justifies holding on to the current paradigm (until proven false). Yet more philosophical work is necessary to clarify what makes a scientific paradigm empirically well supported (over and above entrenchment).

In the meantime, do we have reasons for holding on to the accepted $\Lambda$CDM paradigm, and hence for reconciling the recalcitrant evidence coming from Supernova Ia via Dark Matter and Dark Energy? The answer would seem positive if one considers both the predictive power of General Relativity and the variety of experimental techniques available to measure the equation of state parameter $w$. As always, nature will ultimately answer the question as to whether the current Dark Energy program is on the right track, as Adams and Le Verrier were in postulating Neptune to account for the anomaly in Uranus. In our hands are the experimental and technological tools to search for such an answer.